\documentstyle[12pt]{article}

\begin{document}
\begin{titlepage}

\centerline{\bf ENERGY-MOMENTUM}
\centerline{\bf OF THE GRAVITATIONAL FIELD}
\centerline{\bf IN THE TELEPARALLEL GEOMETRY}
\vskip 1.0cm
\bigskip
\centerline{\it J. W. Maluf$\,^{*}$, J. F. da Rocha-Neto}
\centerline{\it T. M. L. Tor\'{\i}bio and K. H. Castello Branco}
\centerline{\it Instituto de F\'isica}
\centerline{\it Universidade de Bras\'ilia}
\centerline{\it C.P. 04385}
\centerline{\it 70.919-970  Bras\'ilia, DF}  
\centerline{\it Brazil}
\date{}
\begin{abstract}
The Hamiltonian formulation of the teleparallel equivalent
of general relativity (TEGR) without gauge fixing has recently
been established in terms of the Hamiltonian constraint and a
set of six primary constraints. Altogether, they constitute
a set of first class constraints.
In view of the constraint structure we establish
definitions for the energy, momentum and angular momentum
of the gravitational field. In agreement with previous
investigations, the gravitational energy-momentum density
follows from a total divergence that arises in the constraints.
This definition is applied successfully to the calculation
of the irreducible mass of the Kerr black hole. The definition
of the angular momentum of the gravitational field follows from
the integral form of primary constraints that satisfy the
angular momentum algebra.
\end{abstract}
\thispagestyle{empty}
\vfill
\par
\noindent (*) e-mail: wadih@fis.unb.br
\end{titlepage}
\newpage

The dynamics of the gravitational field can be described
in the context of the teleparallel
geometry, where the basic geometrical entity is the tetrad
field $e^a\,_\mu$, ($a$ and $\mu$ are SO(3,1) and
space-time indices, respectively). Teleparallel theories of
gravity are defined on the
Weitzenb\"ock space-time\cite{Weit}, endowed with the affine
connection

$$\Gamma^\lambda_{\mu\nu}=e^{a\lambda} \partial_\mu
e_{a\nu}\;.\eqno(1)$$

\noindent The curvature tensor constructed out of (1) vanishes
identically. This connection defines a space with
teleparallelism, or absolute parallelism\cite{Schouten}.
This geometrical framework was considered by
Einstein\cite{Einstein} in his attempt at unifying gravity and
electromagnetism. 

In the teleparallel geometry it is possible to
establish an alternative description of Einstein's equations.
Such description is given by the teleparallel
equivalent of general relativity
(TEGR)\cite{Mol,Hehl,Hay,Kop,Muller,Nes,Maluf1,Per}.
Gravity  theories in this geometrical framework are
constructed out of the torsion tensor
$T^a\,_{\mu\nu}=\partial_\mu e^a\,_\nu-\partial_\nu e^a\,_\mu$,
which is related to the antisymmetric part of (1).
Interesting features of the TEGR take place in the
Hamiltonian framework. 

The Hamiltonian formulation of the TEGR has been obtained in
Ref. \cite{Maluf1}. In the latter, however, the time gauge was
fixed from the outset. As a consequence of this gauge fixing
the teleparallel geometry is restricted to the
three-dimensional spacelike hypersurface.

In the framework of the TEGR it is possible to make definite
statements about the energy and momentum of the gravitational
field.
A simple expression for the gravitational energy arises in
the Hamiltonian formulation of the TEGR\cite{Maluf1} in
the framework of Schwinger's time gauge
condition\cite{Schwinger}. The energy density is given by a
scalar density in the form of a total divergence that
appears in the Hamiltonian constraint of the
theory\cite{Maluf3}. By applying this definition to several
configurations of the gravitational field
encouraging and satisfactory results have been obtained.
The investigations carried out so far
confirm the consistency and relevance of this energy
expression.

The Hamiltonian formulation of the TEGR without gauge fixing
has recently been established\cite{Maluf2}. Its canonical
structure is different from that obtained in Ref. \cite{Maluf1},
since it is not given in the standard ADM form\cite{ADM}. In
fact it has not been necessary to establish the usual 3+1
decomposition for the metric and tetrad fields. In this
framework we again arrive at an expression for the gravitational
energy, in strict similarity with the procedure adopted in
Ref. \cite{Maluf3}, namely, by interpreting the Hamiltonian
constraint equation as an energy equation for the gravitational
field. Likewise, the gravitational momentum can be defined. 
The gravitational energy-momentum arises as a SO(3,1)
vector. The constraint algebra of the theory suggest that
certain momenta components are related to the gravitational
angular momentum. It turns out to be possible to define, in this
context, the angular momentum of the gravitational field. 

In this paper we investigate the definitions of gravitational
energy and angular momentum that arises in Ref. \cite{Maluf2} in
the framework of the Kerr metric. We recall that the whole
formulation developed in Ref. \cite{Maluf2} is carried out without
enforcing the time gauge condition. It turns out, however, that
consistent values for the gravitational energy are achieved by
requiring the tetrad field to satisfy the time gauge condition.
This amounts to {\it a posteriori} restriction on the
tetrads.\par

\bigskip
\noindent Notation: spacetime indices $\mu, \nu, ...$ and SO(3,1)
indices $a, b, ...$ run from 0 to 3. Time and space indices are
indicated according to $\mu=0,i,\;\;a=(0),(i)$.
The flat, Minkowski spacetime  metric is fixed by
$\eta_{ab}=e_{a\mu} e_{b\nu}g^{\mu\nu}= (-+++)$.\par

\bigskip
\bigskip

The Lagrangian density of the TEGR in empty space-time
is given by\cite{Maluf1,Maluf2}

$$L(e)\;=\;-k\,e\,\biggl( {1\over 4} T^{abc}T_{abc} +
{1\over 2}T^{abc}T_{bac}-T^aT_a\biggr)\;,\eqno(2)$$

\noindent where $k={1\over {16\pi G}}$, $G$ is Newton's
constant, $e=det(e^a\,_\mu)$ and
$T_{abc}=e_b\,^\mu e_c\,^\nu T_{a \mu \nu}$.  
Tetrads transform space-time into SO(3,1) indices
and vice-versa. The trace of the torsion tensor is given by 
$T_b=T^a\,_{ab}\;.$

In the Hamiltonian formulation developed in Ref. \cite{Maluf2}
it has not been made use of any kind of projection of metric
variables to the three-dimensional spacelike hypersurface. The
Hamiltonian was obtained by just rewriting the Lagrangian
density in the form $L=p\dot q -H$. Since there is no time
derivative of $e_{a0}$ in (2), the corresponding momentum
canonically conjugated $\Pi^{a0}$ vanishes identically.
Dispensing with surface terms the total Hamiltonian density
reads\cite{Maluf2} 

$$H(e_{ai},\Pi^{ai})
=H_0+\alpha_{ik}\Gamma^{ik}+\beta_k\Gamma^k\;.\eqno(3)$$

\noindent The Hamiltonian constraint $H_0$ and the primary
constraits $\Gamma^{ik}$ and $\Gamma^k$ are given by,
respectively,

$$H_0\;=\;-e_{a0}\partial_k \Pi^{ak}
-{1\over {4g^{00}}} ke \biggl(g_{ik}g_{jl}P^{ij}P^{kl}-
{1\over 2}P^2\biggr)$$

$$+ke\biggl( {1\over 4}g^{im}g^{nj}T^a\,_{mn}T_{aij}
+{1\over 2}g^{nj}T^i\,_{mn}T^m\,_{ij}
-g^{ik}T^j\,_{ji}T^n\,_{nk}\biggr)\;,\eqno(4)$$

$$\Gamma^{ik}\;=\;-\Gamma^{ki}\;=\;
\Pi^{[ik]}\;-\;k\,e \biggl( -g^{im}g^{kj}T^0\,_{mj}+
(g^{im}g^{0k}-g^{km}g^{0i})T^j\,_{mj} \biggr)
\;,\eqno(5)$$

$$ \Gamma^k \;=\;\Pi^{0k}\; +\;2k\,e\, (
g^{kj}g^{0i}T^0\,_{ij}-g^{0k}g^{0i}T^j\,_{ij}
+g^{00}g^{ik}T^j\,_{ij} )
\;.\eqno(6)$$

\noindent $(..)$ and $\lbrack .. \rbrack$ denote
symmetrization and anti-symmetrization, respectively. In (4)
we have the following definitions

$$P^{ik}\;=\;{1\over {ke}}\Pi^{(ik)}-\Delta^{ik}\;,\eqno(7)$$

$$\Delta^{ik}\;=\;-g^{0m}(
g^{kj}T^i\,_{mj}+g^{ij}T^k\,_{mj}-2g^{ik}T^j\,_{mj})
-(g^{km}g^{0i}+g^{im}g^{0k}) T^j\,_{mj}\;,\eqno(8)$$

\noindent and $P=g_{ik}P^{ik}$. 

It has been shown\cite{Maluf2} that the Lagrange
multipliers $\alpha_{ik}$ and $\beta_{k}$ are determined from
the evolution equations,
$\alpha_{ij}={1\over 2} (T_{i0j}-T_{j0i})\;,$
$\beta_j =T_{00j}\;,$
and that although $e_{a0}$ is present within
the structure of $H_0$, $\Gamma^{ik}$ and $\Gamma^k$, it
actually plays the role of a Lagrange multiplier (see
equation (10) below).

The vanishing of the momentum canonically conjugated to
$e_{a0}$, $\Pi^{a0}$, induces the secondary constraints

$$C^a(x)\equiv {{\delta H}\over {\delta e_{a0}(x)}}=
0\;,\eqno(9)$$

\noindent which satisfy\cite{Maluf2}

$$e_{a0}C^a=H_0\;.\eqno(10)$$

\noindent It is possible to show that $C^a$ may be
written in a simplified form as

$$C^a=e^{a0}H_0+e^{ai}F_i\;,\eqno(11)$$

\noindent where

$$F_i=H_i+\Gamma^m T_{0mi}+\Gamma^{lm}T_{lmi}+
{1\over {2g^{00}}}(g_{ik}g_{jl}P^{kl}-
{1\over 2}g_{ij}P)\Gamma^j\;.\eqno(12)$$

\noindent The vector constraint $H_i$ is given by

$$H_i=-e_{bi}\partial_k \Pi^{bk}-\Pi^{bk} T_{bki}\;.\eqno(13)$$

Therefore if $H_0$ vanishes, $e_{a0}C^a$ also vanishes. Since
$\lbrace e_{a0} \rbrace$ are arbitrary, it follows that
$C^a$ vanishes as well. If in addition we have
$\Gamma^{ij}=\Gamma^j=0$, then we also have $H_i=0$.
Consequently the vanishing of $H_i$ at any instant of time
follows from the vanishing of $H_0$, $\Gamma^{ij}$ and
$\Gamma^j$ at the same instant. Furthermore $H_i$ is
{\it derived} from $H_0$ in the subspace of the phase space
determined by $\Gamma^{ij}=\Gamma^j=0$,

$$e_{ai}{\delta \over {\delta e_{a0}}} H_0=H_i\;.\eqno(14)$$

\bigskip
\bigskip

The Poisson bracket between two
quantites $F$ and $G$ is defined by

$$\lbrace F,G\rbrace=\int d^3x \biggl(
{{\delta F}\over {\delta e_{ai}(x)}}
{{\delta G}\over {\delta\Pi^{ai}(x)}}-
{{\delta F}\over {\delta\Pi^{ai}(x)}}
{{\delta G}\over {\delta e_{ai}(x)}} \biggr)\;,$$

\noindent by means of which we can write down the evolution
equations and evaluate the constraint algebra.

The calculations of the Poisson brackets between the constraints
(4), (5) and (6) are exceedingly complicated. We will just
present the results. The constraint algebra is given by

$$\lbrace H_0(x),H_0(y)\rbrace=0\;,\eqno(15)$$

$$\lbrace \Gamma^i(x),\Gamma^j(y)\rbrace=0\;,\eqno(16)$$

$$\lbrace \Gamma^{ij}(x),\Gamma^{kl}(y)\rbrace={1\over 2}\biggl(
g^{il}\Gamma^{jk}+g^{jk}\Gamma^{il}-
g^{ik}\Gamma^{jl}-g^{jl}\Gamma^{ik}\biggr)\delta(x-y)
\;,\eqno(17)$$

$$\lbrace \Gamma^{ij}(x),\Gamma^k(y)\rbrace =
(g^{0j}\Gamma^{ki}-g^{0i}\Gamma^{kj})\delta(x-y)\;,\eqno(18)$$

$$\lbrace H_0(x),\Gamma^{ij}(y)\rbrace=
\biggl[ {1\over{2g^{00}}}P^{kl}\biggl( {1\over 2}
g_{kl}g_{mn}-g_{km}g_{nl}\biggr)\biggl(
g^{mi}\Gamma^{nj}-g^{mj}\Gamma^{ni}\biggr)+$$

$$+{1\over 2}\biggl(\Gamma^{nj}e^{ai}-\Gamma^{ni}e^{aj}\biggr)
\partial_n e_{a0}\biggr]\delta(x-y)\;,\eqno(19)$$

$$\lbrace H_0(x),\Gamma^i(y)\rbrace =\biggl[g^{0i}H_0+
{1\over {g^{00}}}P^{kl}\biggl({1\over 2}g_{kl}g_{jm}-
g_{kj}g_{ml}\biggr)g^{0j}\Gamma^{mi}$$

$$+\biggl( \Gamma^{ni}e^{a0} 
+\Gamma^n e^{ai}\biggr)  \partial_n e_{a0}+
{1\over 2}\Gamma^{mn}T^i\,_{nm}$$

$$+2\partial_n \Gamma^{ni}+g^{in}\biggl( H_n-\Gamma^jT_{0nj}-
\Gamma^{mj}T_{mnj}\biggr)\biggr]\delta(x-y)$$

$$+\Gamma^{ni}(x)
{\partial \over {{\partial x^n}}}\delta(x-y)\;.\eqno(20)$$

\noindent We note the presence of $H_i$ on the right hand side of
(20). However it poses no problem for the consistency of the
constraints provided $H_0$, $\Gamma^{ik}$ and $\Gamma^k$ are
taken to vanish at the intial time $t=t_0$. Let $\phi(x^i,t)$
represent any of the latter constraints. At the initial time
we have $\phi(x^i,t_0)=0$. At $t_0+\delta t$ we find
$\phi(x^i,t_0+\delta t)=\phi(x^i,t_0)+\dot \phi(x^i,t_0)\delta t$
such that ${\dot \phi(x^i,t_0)}=
\lbrace \phi(x^i,t_0),{\bf H}\rbrace$, where  ${\bf H}$ is the
total Hamiltonian. Since the vanishing of
$H_i$ at an instant of time is a consequence of the vanishing of
$H_0$, $\Gamma^{ik}$ and $\Gamma^k$ at the same time,  the
consistency of the constraints is guaranteed at any $t>t_0$.

One of the main motivations for studying the TEGR 
is that the constraint equations of the theory
can be interpreted as energy-momentum equations. The definition
of the gravitational energy given in Ref. \cite{Maluf3}
was motivated by  interpreting the Hamiltonian constraint
equation $C=0$ of Ref. \cite{Maluf1} as an equation of the
type $C=H-E=0$. A stringent application of such
definition has been made in the context of rotating black
holes\cite{Maluf4}.
In  similarity with Ref. \cite{Maluf3}, in the
present framework again we interpret the $a=0$
component of the constraint equations $C^a=(0)$ as an energy
equation for the gravitational field.

The total divergence that appears in $C^a$ is given by
$-\partial_i \Pi^{ai}$. After implementing the primary
constraints (5) and (6) the momenta $\Pi^{ak}$ reads

$$\Pi^{ak}\;=\;k\,e\biggl\{ 
g^{00}(-g^{kj}T^a\,_{0j}-
e^{aj}T^k\,_{0j}+2e^{ak}T^j\,_{0j})$$

$$+g^{0k}(g^{0j}T^a\,_{0j}+e^{aj}T^0\,_{0j})
\,+e^{a0}(g^{0j}T^k\,_{0j}+g^{kj}T^0\,_{0j})
-2(e^{a0}g^{0k}T^j\,_{0j}+e^{ak}g^{0j}T^0\,_{0j})$$

$$-g^{0i}g^{kj}T^a\,_{ij}+e^{ai}(g^{0j}T^k\,_{ij}-
g^{kj}T^0\,_{ij})-2(g^{0i}e^{ak}-g^{ik}e^{a0})
T^j\,_{ji} \biggr\}\;.\eqno(21)$$

We identify $-\partial_i \Pi^{ai}$, which is the first term
in the expression of $C^a$, as the {\it energy-momentum
density} of the gravitational field. The total energy-momentum
is defined by

$$P^a=-\int_V d^3x \partial_i \Pi^{ai}\;.\eqno(22)$$

\noindent where $V$ is an arbitrary space volume. It is invariant
under coordinate transformations on the three-dimensional
spacelike hypersurface, and transforms
as a vector under the global
SO(3,1) group. The definition above generalizes the
expression previously obtained in Ref. \cite{Maluf3} to tetrad
fields that are not restricted by the time gauge condition.

In analogy with the previous analysis of Ref. 
\cite{Maluf3}, and following M\o ller\cite{Mol}, in the
case of asymptotically flat space-times we may adopt the
boundary conditions

$$e_{a\mu} \simeq \eta_{a\mu}
+ {1\over 2}h_{a\mu}({1\over r})\;,\eqno(23)$$

\noindent in the limit $r \rightarrow \infty$. In the expression
above $\eta_{a\mu}$ is the Minkowski metric tensor and
$h_{a\mu}=h_{\mu a}$ is the first term of the asymptotic
expansion of the metric tensor $g_{\mu \nu}$. Since $h_{a\mu}$
is asymptotically a symmetric tensor, these
boundary conditions impose 6 conditions on the tetrads.
These conditions are not fixed in the body of the theory because
the SO(3,1) is a global (rather than local) symmetry group
(M\o ller called it {\it supplementary conditions}).

Similar conditions
for triads restricted to the three-dimensional spacelike
hypersurface were essential in order to arrive at the ADM
energy\cite{ADM}. In the present approach we also obtain
the ADM energy. Asymptotically flat spacetimes are
defined by (23) together with $\partial_\mu g_{\lambda \nu}
=O({1\over r^2})$, or $\partial_\mu e_{a\nu}=O({1\over r^2})$.
Thus considering the $a=(0)$ component in (22) and integrating
over the whole three-dimensional
spacelike hypersurface we find that all terms of the type
$T^a\,_{0j}$ cancel out, and eventually only the last term in
(21) contributes to the integration. Hence we obtain

$$E\equiv - \int_{V\rightarrow \infty} d^3x \partial_k
\Pi^{(0)k} =-2k\int_{V \rightarrow \infty} d^3x
\partial_k(eg^{ik}e^{(0)0}T^j\,_{ji})$$

$$={1\over {16\pi G}}\int_{S\rightarrow \infty}dS_k(\partial_i
h_{ik}-\partial_k h_{ii}) = E_{ADM}\;.\eqno(24)\;$$

\noindent The energy expression above can be applied to
{\it finite} volumes of space. Therefore it can be used
to obtain the irreducible mass of rotating black holes.
It is the mass of the black hole at the final
stage of Penrose's process of energy extraction, considering
that the maximum possible energy is extracted. It is also
the mass contained within the outer horizon of the black
hole. Every expression
for local or quasi-local gravitational energy must necessarily
yield the value of $M_{irr}$ in the calculation of the
energy contained within the outer event horizon,
since we know
beforehand its value as a function of the initial angular
momentum of the black hole\cite{Chris}.
The evaluation of $M_{irr}$ is a
crucial test for any expression for the gravitational energy
($M_{irr}$ has been obtained by means of different energy
expressions by Bergqvist\cite{Bergqvist}).

In terms of Boyer-Lindquist\cite{BL} coordinates the Kerr metric
tensor is given by

$$ds^2=-{\psi^2 \over\rho^2}dt^2-
{{2\chi sin^2\theta}\over \rho^2}d\phi\,dt
+{\rho^2 \over \Delta}dr^2+
\rho^2 d\theta^2+
{{\Sigma^2 sin^2\theta}\over \rho^2}d\phi^2 \;,\eqno(25)$$

\noindent where $\rho^2=r^2+a^2 cos^2\theta$, 
$\Delta= r^2 +a^2 -2mr$ and

$$\Sigma^2=(r^2+a^2)^2-\Delta a^2\,sin^2\theta\;,$$

$$\psi^2=\Delta -a^2\,sin^2\theta\;,$$

$$\chi=2amr\;. $$

In order to obtain $M_{irr}$ we calculate the $a=(0)$
component of (22) by fixing $V$ to be the volume within the
$r_+=constant$ surface, where $r_+$ is the outer horizon of
the Kerr black hole,

$$r_+=m+\sqrt{m^2-a^2}\;.$$

Of course there is an infinity of tetrads that yield (25),
but only one that leads to a viable expression for $M_{irr}$,
and that gives the correct description of the gravitational
energy. Here we will consider two sets of tetrads. 
The first one is taken to satisfy M\o ller's weak field
approximation (23), and will be denoted by $e^M_{a\mu}$.
It reads

$$e^M_{a\mu}=\pmatrix{
-{\psi \over \rho}\sqrt{1+M^2y^2} & 0 & 0 &
-{{\chi N y}\over {\psi \rho}}sin^2\theta\cr
{{\chi y}\over {\Sigma \rho}}sin\theta\,sin\phi &
{\rho \over \sqrt{\Delta}}sin\theta\,cos\phi &
\rho\, cos\theta\,cos\phi &
-{\Sigma \over \rho}\sqrt{1+M^2N^2y^2}\sin\theta\,sin\phi\cr
-{{\chi y}\over{\Sigma \rho}}sin\theta\, cos\phi &
{\rho \over \sqrt{\Delta}} sin\theta\,sin\phi &
\rho\, cos\theta\,sin\phi &
{\Sigma \over \rho}\sqrt{1+M^2N^2y^2}sin\theta\,cos\phi \cr
0 & {\rho \over \sqrt{\Delta}}cos\theta &
-\rho\, sin\theta & 0 \cr}\,,\eqno(26) $$

\noindent where

$$y^2={{ 2N\sqrt{1+M^2}-(1+N^2)}\over
{4M^2N^2-(1-N^2)^2}}\;,$$

$$M={\chi \over{\Sigma \psi}}sin\theta\;,$$

$$N={{\psi r}\over \Sigma}\;.$$

\noindent The second one satisfies Schwinger's
time gauge condition,

$$e_{(k)}\,^0=e^{(0)}\,_j=0\;,\eqno(27a)$$

\noindent together with the weak field approximation

$$e_{(i)j}\simeq \eta_{ij}+{1\over 2} h_{ij}\;,\eqno(27b)$$

$$h_{ij}=h_{ji}\;.\eqno(27c)$$

\noindent It is given by

$$e^S_{a\mu}=\pmatrix{
-{1\over \rho}\sqrt{\psi^2+{\chi^2\over \Sigma^2}sin^2\theta} &
0&0&0\cr
{\chi \over {\Sigma \rho}}sin\theta\,sin\phi &
{\rho \over \sqrt{\Delta}}sin\theta\,cos\phi &
\rho\,cos\theta\,cos\phi &
-{\Sigma \over \rho} sin\theta\,sin\phi\cr
-{\chi \over{\Sigma \rho}}sin\theta\,cos\phi &
{\rho \over \sqrt{\Delta}}sin\theta\,sin\phi &
\rho\,cos\theta\,sin\phi &
{\Sigma \over \rho}sin\theta\,cos\phi\cr
0&{\rho \over \sqrt{\Delta}}cos\theta&-\rho\,sin\theta&0\cr}
\,.\eqno(28)$$

Considering (21) and (22) we
calculate the gravitational energy contained within
a surface enclosing the black hole, determined by a
constant radius $r_0$, and then take the limit
$r_0\rightarrow r_+$, both for $e^M_{a\mu}$ and for
$e^S_{a\mu}$. The final expression arises as a function
of the angular momentum per unit mass $a$.


By using (28) we find that the resulting
energy expression, 
$E\lbrack e^S_{a\mu}\rbrack$,  is precisely the same one
obtained in Ref. \cite{Maluf4}, and therefore it agrees
remarkably well with the expression of $2M_{irr}$ as a
function of $a$. It reads

$$E\lbrack e^S_{a\mu}\rbrack =
m\biggl[ {\sqrt{2p}\over 4}+{{6p-k^2}\over 4k} ln
\biggl( {{\sqrt{2p} +k}\over p} \biggr) \biggr]\;,\eqno(29)$$

\noindent where

$$p=1+\sqrt{1-k^2}\;\;\;\;,\;\;\;\;a=km, 0\leq k \leq 1\;.$$

The energy expression $E\lbrack e^M_{a\mu}\rbrack$
that is obtained by considering (26)
deviates from $M_{irr}$. For values of $a \simeq 0.8m$ we
find that $E\lbrack e^M_{a\mu}\rbrack >2m$. However this
value must be always less than or equal to $2m$. The expression
is given by

$$E\lbrack e^M_{a\mu}\rbrack = {m\over 4}
\int_0^\pi d\theta\,sin\theta\,\bigg[
\sqrt{p^2+k^2 cos\theta}+{{py}\over \sqrt{p^2+k^2 cos\theta}}$$

$$+{ {2p^3y} \over{(p^2+k^2 cos\theta)^{3\over 2}}}-
{ {y(p-1) \sqrt{p^2+k^2 cos\theta}}\over 2}\biggr]\;.\eqno(30)$$

\noindent Details will
be presented elsewhere\cite{Maluf5}.

We will ascribe generality to the result above and
assume that tetrads satisfying the time gauge
condition (27a), together with (27b,c), yield the correct
description of the gravitational energy.\par

\bigskip

The Poisson bracket (17) strongly suggestes that $\Gamma^{ik}$
is related to the angular momentum of the gravitational field.
In similarity with the definition of the gravitational
energy-momentum, we assume that the gravitational angular
momentum $M^{ik}$ is obtained from the integral form of the
constraint equation $\Gamma^{ik}=0$. Therefore we define

$$M^{ik}=\int d^3x\, \Pi^{[ik]}$$

$$
= \int d^3x\, k\,e \biggl[ -g^{im}g^{kj}T^0\,_{mj}+
(g^{im}g^{0k}-g^{km}g^{0i})T^j\,_{mj} \biggr]
\;.\eqno(31)$$

Considering the Kerr space-time,
the evaluation of $M^{ik}$ out of (28) involves the evaluation
of very intricate integrals. This issue is currently under
investigation\cite{Maluf5}. Here we just mention that  the
application of (31) to the simple case of the 
metric associated to a thin, slowly
rotating mass shell as described by Cohen\cite{Cohen} yields
the Newtonian expression for the angular momentum of the
source. This result is obtained by integrating
(31) over the {\it whole} three-dimensional space, and making
use of the time gauge condition (27).
It must be noted that such metric tensor
corresponds to the asymptotic form of Kerr's metric tensor
in the limit of small angular momentum. \par

\bigskip

Although the time gauge condition is clearly important in
analysis of the Hamiltonian formulation of tetrad type theories
of gravity, its relevance in the context of the present
investigation is not fully understood. The analysis of Refs. 
\cite{Maluf1,Maluf3} was developed under the assumption of the time
gauge condition, and as a consequence the teleparallel
geometry was restricted to the three-dimensional spacelike
hypersurface. By not assuming any {\it a priori} restriction
on the tetrads the teleparallel geometry is extended to the
four-dimensional space-time, and yet the time gauge condition
continues to play a special role in the description of the
energy of the gravitational field. The question regarding the
relevance of the time gauge condition in the present analysis
must be further investigated.\par
\bigskip
\noindent {\it Acknowledgments}
\noindent J. F. da R. N., T. M. L. T. and K. H. C. B. are grateful
to the Brazilian Agency CAPES for financial support.\par

\end{document}